\documentclass{ws-mpla}
\usepackage{graphics}
\usepackage{graphicx}
\usepackage{amsmath}
\usepackage{amsfonts}
\usepackage{amssymb}
\newcommand{\p}{\partial}
\newcommand{\ie}{\emph{i.e., }}
\newcommand{\reff}[1]{(\ref{#1})}
\newcommand{\eref}[1]{Eq.\reff{#1}}

\newcommand{\figref}[1]{FIG.\ref{#1}}
\newcommand{\ub}{\bar{u}}
\newcommand{\vb}{\bar{v}}
\usepackage{color}

\begin{document}

\markboth{R. Martorelli, G. Montani and N. Carlevaro}
{Effect of a stochastic electric field on plasma confinement in FTU}


\catchline{}{}{}{}{}

%%%%%%%%%%%%%%%%%%%%%%%%%%%%%%%%%%%%%%%%%%%%%%%%%%%%%%%%%%%%%%%%%%%%

\title{EFFECT OF A STOCHASTIC ELECTRIC FIELD\\ ON PLASMA CONFINEMENT IN FTU}

\author{\footnotesize ROBERTO MARTORELLI}
\address{Institut fuer Theoretische Physik I, Universitaet Duesseldorf, Dusseldorf 40225, Germany.}

\author{\footnotesize GIOVANNI MONTANI}
\address{ENEA - R.C. Frascati, Via E. Fermi, 45 (00044) Frascati (Roma), Italy;\\
Physics Department, ``Sapienza'' University of Rome, P.le Aldo Moro 5, (00185) Roma, Italy.}

\author{\footnotesize NAKIA CARLEVARO}
\address{ENEA - R.C. Frascati, Via E. Fermi, 45 (00044) Frascati (Roma), Italy.}

\maketitle

\begin{abstract}
We discuss a stochastic model for the behavior of electrons in a magnetically confined plasma having axial symmetry. The aim of the work is to provide an explanation for the density limit observed in the Frascati Tokamak Upgrade machine. The dynamical framework deals with an electron embedded in a stationary and uniform magnetic field and affected by an orthogonal random electric field. The behavior of the average plasma profile is determined by the appropriate Fokker-Planck equation associated to the considered model and the disruptive effects of the stochastic electric field is shown. The comparison between the addressed model and the experimental data allows to fix the relevant spatial scale of such a stochastic field. It is found to be of the order of the Tokamak micro-physics scale, \ie few millimeters. Moreover, it is clarified how the diffusion process outlines a dependence on the magnetic field as $\sim B^{-3/2}$.
\end{abstract}

\ccode{PACS No.: 52.35.-g; 05.10.Gg; 52.20.Dq}

\section{Introduction}
\label{Introduction}
A long standing problem related to fusion plasmas consists of reaching the requested density to trigger the ignition. Experiments can now regularly operate at densities of the order of $10^{20}$ m$^{-3}$ but, according to the Lawson criterion \cite{Wesson}, it is also necessary to maintain this regime for a specific confinement time. This is clearly a challenging task, especially because increasing the density gives rise to instabilities often leading to the disruption of the plasma. In this work, we analyze the intrinsic dependence of the central density (and its possible limit) in function of the toroidal magnetic field, as recently observed in the Frascati Tokamak Upgrade (FTU) experiment. In Ref.~\refcite{Pucella}, a distinction between the central line averaged density and the edge density has been proposed. In particular, while a Greenwald-like limit \cite{Greenwald} still holds for the latter, the former shows a maximum which scales with the toroidal magnetic field amplitude. This scaling law is determined by experimental observations and, for the FTU case, reads as $n_{FTU}=0.19\times10^{20}\,B^{3/2}\;\,\mathrm{m}^{-3}\mathrm{T}^{-3/2}$, where the toroidal magnetic field $B$ is in Tesla units. It is natural to argue that increasing the fuel would unavoidably bring to an expansion of the inner part of the plasma and, in turn, to the crossing of the Greenwald-limit as a consequence of the profile spread. By other words, it is relevant to recall how the central line limit could be an effect induced by the plasma edge and not necessarily an intrinsic property of the plasma, as we are going to infer. In fact, it is still not clear if the actual dependence of the central density limit on $B$ is a physical feature or just a technical limitation. In Ref.~\refcite{Carati}, it has been suggested that such a dependence could be explained through the perturbative electric field generated by the charge fluctuations in the plasma. This way, the following behavior, disagreeing with the FTU experiment, is provided: $n=3\,\epsilon_0B^2/2m=0.14\times10^{20}\,B^{2}\;\,\mathrm{m}^{-3}\mathrm{T}^{-2}$, where $\epsilon_0$ and $m$ are the vacuum dielectric constant and the electron mass in SI units, respectively. This result is analogous to the Brillouin limit for non-neutral plasma \cite{Brillouin}. Furthermore, plasma disruption due to stochasticity effects has been also studied in Refs.~\refcite{Gates,Boozer}.

The main goal of this paper is to show how the FTU measurements on the central density behavior (as function of the toroidal field) can be interpreted in terms of the effect of a stochastic electric field, deconfining the electrons from their Larmor orbits. Specifically, we determine the precise spatial scale of such a field, as request to reproduce the FTU data in the field-density plane and it turns out to be of the millimeter order. This suggests that the disruption process, in correspondence of high central density, can be triggered by the electrostatic turbulence at the microphysical Tokamak scales, like the drift wave turbulence. The technical paradigm to achieve this goal is based on the formulation of a Fokker-Planck equation in the velocity space as result of a Gaussian noise in the force acting on the electrons. The emerging electron profile contains a free parameter (in the form of a characteristic time) which, once fixed by fitting the FTU data, yields the electric turbulence scale. Indeed, this scenario must be matched with the triggering of a Greenwald instability, as result of the flattening of a very dense central region of the plasma. In this sense, the observed FTU behavior can be interpreted in terms of two overlapping phenomena: on one hand, as here clarified, the stochastic electric force affects the Larmor motion, on the other hand, associated flattening of the density profile triggers the plasma-wall interaction.

The paper is organized as follows. In the second Section, we introduce the analytical model for the plasma electron dynamics in the presence of an external magnetic field and a perturbative electric field. By solving the Fokker-Planck equation related to the dynamics, we are able to obtain an expression for the plasma density in terms of the perturbation scale. In the third Section, we apply the solution of the analytical model to the FTU configuration and, by properly characterizing the plasma disruption, we are able to determine the spatial scale of the electric perturbation responsible for the density limit. Concluding remarks follows.

\section{Stochastic model}
\label{Stochastic model}
We here develop an analytical model which takes account for the presence of a stochastic electric field affecting the motion of the electrons around the magnetic field line. The goal of the present study results in determining the spatial scale of the electric-field perturbations in order to recover the proper density limit function as observed in the FTU experiment.

The addressed system is composed by a one-component plasma with a neutralizing background, subjected to a magnetic field $B$ directed along the $z$ axis (($x,\,y,\,z$) being Cartesian coordinates), \ie $\textbf{B}=B\hat{\textbf{e}}_z$, modeling the toroidal magnetic field of a Tokamak device. A transverse (lying on the $x$-$y$ plane) stochastic electric field of amplitude $E(t)$ is also considered. In spite of the simplicity of the model, this is a good approximation for a Tokamak configuration because of the local character of the effects we are dealing with. In fact, we are interested in investigating how the disruption can be a consequence of a stochastic motion of the electrons, no longer confined in the Larmor orbits. Longitudinal electric fields can clearly exist in the plasma, but they are mainly responsible for the electron motion along the magnetic field lines, almost non-affecting the Larmor dynamics. Furthermore, as far as the bulk physics of the plasma is well represented by an ideal magneto-hydrodynamical scenario (like it is observed in many Tokamak quasi-stationary configurations), such longitudinal electric fields must essentially vanish. Finally, it is worth noting that, in our approach, we are neglecting all other transverse electric contributions which do not resemble a stochastic profile, because they are expected to provide a net measurable effect. Since we are retaining only basic features of the configuration, the present results could be extended also to more general plasma sheets, like for instance the Stellarator profile, at the price of a better characterization of the magnetic field line structure. The dynamics of an electron, moving in the configuration described above, can be expressed via the following system of equations:
\begin{subequations}\label{eq.4}
\begin{align}
\dot{x}=&\,\,u\;,\\
\dot{y}=&\,\, v\;,\\
\dot{z}=&\,\, w\;,\\
\dot{u}=&\,\, \omega_cv+eE(t)/m\;,\\
\dot{v}=&\,\, -\omega_cu+eE(t)/m\;,\\
\dot{w}=&\,\, 0\;,
\end{align}
\end{subequations}
where $\omega_c=eB/m$ is the cyclotron frequency ($e$ being the electron charge). Due to the stochastic nature of the electric field, it turns out necessary to characterize its fluctuations via a statistical distribution. We assign here a Gaussian noise, defined by the conditions
\begin{subequations}\label{eq.5}
\begin{align}
\langle E_i(t)\rangle&=0\;,\\
\langle E_i(t)E_j(t')\rangle&=\eta\;\delta_{ij}\delta(t-t')\;,
\end{align}
\end{subequations}
where $\eta$ is the diffusion coefficient, while $i$ and $j$ labels the electric field components. The expression of the diffusion coefficient can be found in Ref.~\refcite{Iglesias}, but associated to a stationary Gaussian electric field. In this respect, we assume here that the result of Ref.~\refcite{Iglesias} can be directly applied to our configuration once defined a time scale $\tau$ for which the electric field can be considered as stationary. This leads to the following expression of the diffusion coefficient:
\begin{equation}\label{eq.6}
\eta=n\tau k_BT/\epsilon_0\;,
\end{equation}
where $n$ is the electron density in $\mathrm{m}^{-3}$, $k_B$ the Boltzmann constant, $T$ the temperature in $K$.

Since our study is aimed to fix the spatial scale of the electric field perturbations responsible for the scaling law observed in FTU, we need to express $\tau$ as function of the plasma density. To this end, we solve the dynamical system \reff{eq.4} by addressing the associated Fokker-Planck equation:
\begin{align}\label{eq.7}
\p_t f = -u\p_x f - v\p_y f - w\p_z f 
-\omega_c v\p_u f + \omega_c u\p_v f + \eta(\p^2_u f + \p^2_v f)\;,
\end{align}
%\begin{equation}
%\label{eq.7}
%\frac{\partial f}{\partial t}=-u\frac{\partial f}{\partial x}-v\frac{\partial f}{\partial y}-w\frac{\partial f}{\partial z}
%-\omega_cv\frac{\partial f}{\partial u}+\omega_cu\frac{\partial f}{\partial v}+D(\frac{\partial^2 f}{\partial u^2}
%+\frac{\partial^2 f}{\partial v^2})\;,
%\end{equation}
where $f=f(t,x,y,z,u,v,w)$ is the probability distribution function for the electrons in the phase space. It is worth noting that the dynamics on the $(x,y)$ plane does not affect the $z$ component of the motion since no force acts in that direction and the value of $w$ is a constant of motion fixed by the initial conditions. Thus, we can factorize the distribution function as the product $f(t,x,y,z,u,v,w)=g(t,x,y,u,v)h(z,w)$ and apply the variable separation procedure. Henceforth, we will analyze the dynamics of the function $g$ only.

Since we have to focus our analysis on the transverse dynamics, we set the following coordinate transformation \cite{Jimenez} in order to properly face the coupling term arising from the Lorentz force:
\begin{subequations}\label{eq.8}
\begin{eqnarray}
\ub=u\cos(\omega_ct)-v\sin(\omega_ct)\;,\\
\vb=u\sin(\omega_ct)+v\cos(\omega_ct)\;.
\end{eqnarray}
\end{subequations}
Since such a transformation has the determinant of the Jacobian equal to one (\ie a rotation in the velocity space), it does not affect the statistical properties of the electric field and does not add any contribution to the integral measure. Implementing a Fourier transformation on the spatial coordinate in \eref{eq.7}, we obtain the following Fokker-Planck equation
\begin{align}
&\p_t\tilde{g}=-ik_x\tilde{g}[\ub\cos(\omega_c t)+\vb\sin(\omega_c t)]+\nonumber\\
&\quad+ik_y\tilde{g}[\ub\sin(\omega_c t)-\vb\cos(\omega_c t)]
+\eta(\p^2_{\ub}\tilde{g}+\p^2_{\vb}\tilde{g})\;,
\label{eq.9}
\end{align}
where $\tilde{g}=\tilde{g}(t,k_x,k_y,\ub,\vb)$ is the Fourier transform of the $g$-distribution. In order to solve this partial differential equation having time-dependent coefficients, we adopt the Lie algebraic methods \cite{Wei}. Defining the initial condition for the distribution function as $\tilde{g}_0=\tilde{g}(0,k_x,k_y,\ub,\vb)$, we obtain
\begin{align}\label{eq.10}
&\tilde{g}=\frac{1}{4\pi\eta t}\;
\exp\!\Big[-\frac{i}{\omega_c}(\ub\Gamma-\vb\Sigma)-\frac{\eta}{\omega_c^2}(k_x^2+k_y^2)t\Big]\times\\
&\times\!\!\!\int\!\!\!\!\int_{-\infty}^{+\infty}\!\!\!\!\!\!\!\exp\!\Big[\frac{(\vb+\alpha\Gamma-y)^2+(\ub-\alpha\Sigma-y')^2}{-4\eta t}\Big]\tilde{g}_0\;dydy'\;,\nonumber
\end{align}
where
\begin{subequations}\label{eq.11}
\begin{align}
&\Gamma=k_x\sin(\omega_ct)+k_y\cos(\omega_ct)\;,\\
&\Sigma=k_y\sin(\omega_ct)-k_x\cos(\omega_ct)\;,\\
&\alpha=2i\eta/\omega_c^2\;.
\end{align}
\end{subequations}

The complete solution must rely on the explicit form of the initial condition $\tilde{g}_0(k_x,k_y,\ub,\vb$). Since we are interested in determining the density threshold for the electron deconfinement, a suitable choice for the initial condition is the density profile corresponding to a quasi-stable configuration of the plasma. This way, we can investigate the proper conditions for which the model exhibits a disruption. In the real space, an example of such an initial profile is shown in Ref.~\refcite{Tudisco} and can be reasonably approximated with a Gaussian of both space and velocity variables. Thus, the function $\tilde{g}_0$ reads as
\begin{equation}
\label{eq.12}
\tilde{g}_0(k_x,k_y,\ub,\vb)=\frac{m}{2\pi k_BT}\frac{1}{2\pi \sigma^2}
\;e^{-\frac{m\ub^{2}+m\vb^{2}}{2 k_B T}-\frac{k_x^2+k_y^2}{2\sigma^2}}\;,
\end{equation}
where $\sigma$ is the variance in the $k$-space.

The physical phenomena we are going to describe concern the inner part of the confined plasma as shown in Ref.~\refcite{Pucella}. Therefore, we reasonably set $1/\sigma$ equal to one third of the minor radius of the FTU Tokamak device, \ie $\sigma=8.33$m$^{-1}$. Moreover, being focused on the role played by the plasma density, the temperature is here and in the following considered as a fixed parameter. However, it is well known that the temperature decreases in correspondence to the increase of the radius of the plasma configuration. Nonetheless, such a phenomenon is clearly a consequence of the disruption onset and, in first approximation especially in the inner part of the plasma column, it can be neglected when searching for the electron deconfinement mechanism. Furthermore, as we shall see in the end of the paper, the crucial quantities entering the addressed mechanism are the Larmor radius and the Debye length and their ratio is temperature independent. Substituting the initial condition \eref{eq.12} in the integral of \eref{eq.10}, we obtain the Fourier transformed distribution function for the electrons. In order to get a pure spatial representation of the system, we can integrate the obtained distribution function over the velocity space. Transforming back to the real space by exploiting the cylindrical symmetry of the addressed scheme, we reach the following distribution function for plasma electrons in terms of the time $t$ and the cylindrical radial coordinate $r$:
\begin{align}
g(t,r)=\frac{\sigma^2B^2e^2\epsilon_0r}{\sigma^2k_BT(\epsilon_0m+4e^2n\tau t)+B^2e^2\epsilon_0}\times\qquad\nonumber\\
\times\exp\Big[-\frac{\sigma^2B^2e^2\epsilon_0r^2}{2[B^2e^2\epsilon_0+\sigma^2kT(\epsilon_0m+4e^2n\tau t)]}\Big]\;.
\label{eq.13}
\end{align}
\begin{figure}
\centering\includegraphics[width=0.6\textwidth]{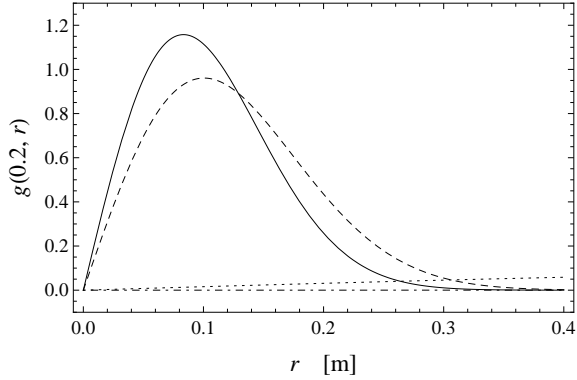}
\caption{\label{fig.1} Radial dependence (at fixed time $t=0.2$s) of the distribution function \reff{eq.13} for different values of the plasma density: $n=10^5$m$^{-3}$ (normal line); $n=10^{10}$m$^{-3}$ (dashed); $n=10^{15}$m$^{-3}$ (dotted); $n=10^{20}$m$^{-3}$ (dot-dashed). The physical parameters are set as: $B=8$T, $T=2$keV and $\tau=\omega_p^{-1}$ with the plasma frequency $\omega_p$ defined as $\omega_p=\sqrt{ne^2/\epsilon_0m}$ (this is the natural time scale for the electron cloud fluctuations).}
\end{figure}

A first insight about the plasma behavior can be traced by studying the radial dependence of the distribution function for different plasma density values, qualitatively represented in \figref{fig.1} for a sufficiently long time ($t=0.2$s) and a typical $\tau$ value. Increasing $n$, the distribution function spreads out underlying a loss of localization for the electrons, triggering the onset of the disruption. It is worth noting that the distribution function becomes extremely flat when the density is approaching the observed density limit, \ie $n\simeq10^{20}$m$^{-3}$.

\section{Prediction of the Model}\label{premod}
Let us now face the characterization of the spatial scale for the perturbing electric field. This can be reached by studying the evolution of the average radius, which can be evaluated by the distribution function as $\langle r(t)\rangle=\int g(t,r)rdr$:
\begin{equation}\label{eq.14}
\langle r(t)\rangle=\sqrt{\frac{\pi mk_BT}{2e^2B^2}\Big(1+\frac{4ne^2\tau}{m\epsilon_0}t\Big)+\frac{\pi}{2\sigma^2}}\;.
\end{equation}
Since we are dealing with a stochastic process, a meaningful description is obtained by studying the variance $\langle r(t)^2\rangle-\langle r(t)\rangle^2$: large values of the variance imply that the electrons tend to drift away from the mean value, indicating the onset of the disruption process. Analyzing the ratio between the variance and the mean value, it can be shown how such a ratio remains below the unity for a wide range of the model parameters. Thus, we are let to perform our study focusing attention to the mean value only.

\begin{figure}
\centering\includegraphics[width=0.6\textwidth]{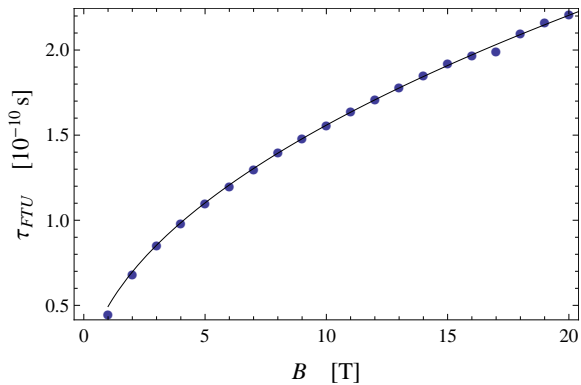}
\caption{\label{fig.2}Time scale $\tau$ (in seconds) as a function of the magnetic field $B$ (in Tesla). The dots are the data sampled according to \eref{eq.15} while the line represent the function that best fits the data.}
\end{figure}

In order to obtain the correct scale of the electric perturbation, we firstly substitute the density $n$ in \eref{eq.14} with the phenomenological FTU scaling law defined as $n_{FTU}=0.19\times10^{20}\,B^{3/2}\;\,\mathrm{m}^{-3}\mathrm{T}^{-3/2}$. This way, we obtain the following expression for the mean value of the radius:
\begin{equation}\label{eq.15}
\langle r(t)\rangle=1.25\Big(\frac{B^2e^2+\sigma^2k_BTm}{\sigma^2B^2e^2}+
\frac{7.6\times10^{19}k_BT\tau t}{\epsilon_0 \sqrt{B}}\Big)^{1/2}\;,
\end{equation}
where all the variables are expressed in SI units. It is now possible to sample the values of $\tau$ yielding to the loss of confinement as the external magnetic field varies. In this sense, a natural prescription consists of marking when the mean value of the radius reaches the edge of the Tokamak. Moreover, it has to be set a time scale for this event to occur, since, in this configuration, the electrons are always subjected to a transverse drift. We chose such a scale as the inverse of the plasma frequency $\omega_p$, since this is a typical deconfinement time for the electrons. By other words, we are physically stating that the electron drift results to be faster than the restoring force due to the ions. The parameters used for the $\tau$-sampling are those of FTU \cite{Pucella}, \ie $\langle r (1/\omega_p)\rangle=0.31$m and $T=2$keV, and, in \figref{fig.2}, we present the values for $\tau$ and the best data fit found as $\tau_{FTU}=5\times10^{-11} B^{1/2}$.
Since we are interested in the density dependence of the parameter $\tau$, let us now use the expression of $n_{FTU}$, obtaining (with obvious notation)
\begin{equation}\label{eq.16}
\tau_{FTU}=1.8\times10^{-17}n^{1/3}\;.
\end{equation}

This behavior sets the scale of the electric field fluctuations required to account for the disruption processed observed in Ref.~\refcite{Pucella}. It can be argued how the parameter $\tau_{FTU}$ goes like the inverse of the particle distance and it is now possible to evaluate the spatial scale of the perturbations by implementing the proper parameters of FTU introduce above. By considering the critical density $n=10^{21}$m$^{-3}$, we get
\begin{equation}\label{eq.17}
L=v_T\ \tau_{FTU}\simeq0.002\ \mathrm{m}\;,
\end{equation}
where $v_T=\sqrt{k_BT/m}\simeq6\times10^{6}$m/s is the thermal velocity. This way, we have reached a complete description of the stochastic electric field responsible for the plasma disruption. The space scale results to be of the order of the millimeter, \ie the microscopic scale for the plasma. This states that an interaction between different scales in the Tokamak emerges, since a microscopic perturbation affects the dynamics of the confined electron which takes place on the Larmor scale. Therefore, we here infer that the phenomena responsible for the disruption have to be identified in the electrostatic turbulence on the millimeter scale.

The present analysis implicitly relies on the presence of a diffusion coefficient. In order to compare this scheme with other approach present in the literature (see for instance Refs.~\refcite{B49,S60}) let us now extrapolate the spatial diffusion coefficient as $\eta_r=\langle r^2\rangle/t$. Using the distribution function \reff{eq.13} and the fitting function of $\tau_{FTU}$, one easily obtains
\begin{equation}\label{eq.20}
\eta_r=\bar{\eta}_r(t)+\eta_r^{(B)}\;,\qquad
\bar{\eta}_r(t)=\frac{2}{\sigma^2t}+\frac{2k_BTm}{e^2B^2t}\;,\quad
\eta_r^{(B)}=\frac{4\times10^{-10} nk_BT}{\epsilon_0 B^{3/2}}\;.
\end{equation}
\begin{figure}
\centering\includegraphics[width=226pt]{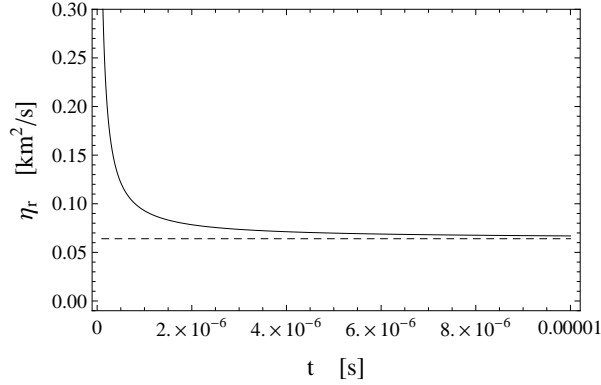}
\caption{\label{fig.3} Temporal behavior of the spatial diffusion coefficient $\eta_r$ described in \eref{eq.20}. The dashed line represents the numerical value of $\eta_r^{(B)}$. The physical quantities are set as proper of FTU.}
\end{figure}
In the expression above, the leading term is $\eta_r^{(B)}$, since $\bar{\eta}_r$ decrease in time. In particular, in the parameter range used in this work, it can be found $\bar{\eta}_r/\eta_r^{(B)}\simeq10^{-2}$ for $t\simeq5\times10^{-5}$s (the complete behavior of $\eta_r(t)$ is plotted in \figref{fig.3} together with the value of $\eta_r^{(B)}$). It is worth noting that, since $\tau$ is the time duration of a single realization of the stochastic electric field, the time scale on which the diffusion coefficient must be evaluated has to be much grater than $\tau$ (our example is indeed implemented for $t\simeq10^{-5}\mathrm{s}\simeq10^{5}\tau_{FTU}$). Thus, we can infer that the density limit in FTU is due to a diffusion coefficient which behaves like $B^{-3/2}$. Such a diffusive profile as function of the magnetic field has been already discussed in Ref.~\refcite{Hsu}. This last result finds its justification in the similarity between the hypothesis used in Ref.~\refcite{Hsu} and those ones underlying our paradigm, \ie mainly the finiteness of the Larmor radius and its distortion by means of a stochastic deconfinig force. 

We observe that, in a Tokamak plasma, the electrostatic perturbations exist almost regardless the density profile (as emerging from \eref{eq.16}), while observations show how the disruption occurs only when the density parameter reaches a specific threshold. A possible explanation for such a discrepancy can be found by comparing the Debye length $\lambda_D=\sqrt{\epsilon_0k_BT/e^2n}$ with respect to the Larmor radius $r_L=v_T m/eB$, both evaluated at the disruption conditions. Considering, as above, the FTU configuration, we obtain the following relation:
\begin{equation}
\label{eq.18}
\lambda_D\simeq r_L\simeq10^{-5}\mathrm{m}\;,
\end{equation}
implying that the trapped electrons move on orbits that overlap the Debye shield. The relation above is evaluated by using the critical density value, and, further increasing the density would imply $r_L\gtrsim\lambda_D$. By other words, the disruption takes place when the electrostatic fluctuations, felt by the Larmor orbits, are no longer shielded by the charge readjustment within the Debye scale. This means that the description of the disruption phenomenon requires a Finite Larmor Radius analysis, the same approach used in Ref.~\refcite{Hsu}.

\section{Conclusions}\label{Conclusion}
We have proposed a basic model to explain the density versus magnetic field scaling law observed in the FTU device \cite{Pucella} by means of the presence of electric field stochastic fluctuations. We have analyzed the dynamics of an electron subjected to an external magnetic field (which mimics the Tokamak toroidal one), and a transverse stochastic electric field reproducing the electrostatic turbulence profile, mainly due to non-linear drift phenomena. We have found the proper microscopic value of the electrostatic turbulence requested to reproduce the FTU phenomenology. The relevant stochastic perturbation is found to occur at the typical microscopic scale of the Tokamak plasma, \ie few milliliters. As an explanation of such a phenomenological issue, we have inferred that the Debye shield on the electron loses its efficiency near the critical density value.


\begin{thebibliography}{00}

\bibitem{Wesson}
J. Wesson, \emph{Tokamaks} (Clarendon Press, Oxford, 1997).

\bibitem{Pucella}
G. Pucella et al., \emph{Nucl. Fus.} \textbf{53}, 023007 (2013).

\bibitem{Greenwald}
M. Greenwald, \emph{Pl. Phys. Contr. Fus.} \textbf{44}, R27 (2002).

\bibitem{Carati}
A. Carati et al., \emph{Chaos} \textbf{22}, 033124 (2012).

\bibitem{Brillouin}
L. Brillouin, \emph{Phys. Rev.} \textbf{67}, 260 (1945).

\bibitem{Gates}
D.A. Gates, L. Delgado-Aparicio, \emph{Phys. Rev. Lett.} \textbf{43}, 267 (2012).

\bibitem{Boozer}
A.H. Boozer, \emph{Phys. Plasmas} \textbf{12}, 104502 (2005).

\bibitem{Iglesias}
C. Iglesias, J. Lebowitz, D. MacGowan, \emph{Phys. Rev. A} \textbf{28}, 1667 (1983).

\bibitem{Jimenez}
J. Jimenez-Aquino, M. Romero-Bastida, \emph{Phys. Rev. E} \textbf{76}, 021106 (2007).

\bibitem{Wei}
J. Wei, E. Norman, \emph{Proc. Amer. Math. Soc.} \textbf{15}, 2034065 (1964).

\bibitem{Tudisco}
O. Tudisco et al., \emph{Fus. Eng. Des.} \textbf{85}, 902 (2010).

\bibitem{B49}
D. Bohm, \emph{The characteristics of electrical discharges in magnetic fields}, Eds. A. Guthrie and R.K. Wakerling (MacGraw-Hill, New York, 1949).

\bibitem{S60}
L. Spitzer, \emph{Phys. Fluids} \textbf{3}(4), 659 (1960).

\bibitem{Hsu}
J.-Y. Hsu et al., \emph{Phys. Plasmas} \textbf{20}, 062302 (2013).

\end{thebibliography}
\end{document}